\documentclass[aps, prl, reprint, superscriptaddress,floatfix,amsmath,amssymb]{revtex4-1}

\usepackage{amssymb}
\usepackage{amsmath}
\usepackage[dvips]{graphicx}
\usepackage{dcolumn}
\usepackage{bm}
\usepackage{natbib}

\usepackage[usenames,dvipsnames,svgnames,table]{xcolor}

\newcommand{\CuSe}{{\text{Cu}_2 \text{O} \text{Se} \text{O}_3}}

\newcommand{\no}{\noindent}

\renewcommand{\vec}[1]{\mathbf{#1}}
\newcommand{\ra}{{\vec r}}
\newcommand{\ka}{{\vec k}}

\newcommand{\bra}[1]{ \langle #1 |}
\newcommand{\ket}[1]{|#1  \rangle}

\begin{document}

\title{Direct control of the skyrmion phase stability by electric field in a magnetoelectric insulator}


\affiliation{Laboratory for Quantum Magnetism  (LQM), {\'E}cole Polytechnique F\'{e}d\'{e}rale de Lausanne (EPFL), \\ CH-1015 Lausanne, Switzerland}
\affiliation{Laboratory for Neutron Scattering and Imaging  (LNS), Paul Scherrer Institut (PSI), \\ CH-5232 Villigen, Switzerland}
\affiliation{Laboratory for  Scientific Developments and Novel Materials (LDM), Paul Scherrer Institut (PSI), \\ CH-5232 Villigen, Switzerland}
\affiliation{Crystal Growth Facility, {\'E}cole Polytechnique F\'{e}d\'{e}rale de Lausanne (EPFL), \\
CH-1015 Lausanne, Switzerland}


\author{A.\,J.\, Kruchkov}
\email{alex.kruchkov@epfl.ch}
\affiliation{Laboratory for Quantum Magnetism (LQM), {\'E}cole Polytechnique F\'{e}d\'{e}rale de Lausanne (EPFL), \\ CH-1015 Lausanne, Switzerland}
\author{J.\,S.\,White}
\affiliation{Laboratory for Neutron Scattering and Imaging (LNS), Paul Scherrer Institut (PSI), \\ CH-5232 Villigen, Switzerland}
\author{M.\,Bartkowiak}
\affiliation{Laboratory for Scientific Developments and Novel Materials (LDM), Paul Scherrer Institut (PSI), \\ CH-5232 Villigen, Switzerland}
\author{I.\, Zivcovic}
\affiliation{Laboratory for Quantum Magnetism  (LQM), {\'E}cole Polytechnique F\'{e}d\'{e}rale de Lausanne (EPFL), \\ CH-1015 Lausanne, Switzerland}
\author{A.\,Magrez}
\affiliation{Crystal Growth Facility, {\'E}cole Polytechnique F\'{e}d\'{e}rale de Lausanne (EPFL), \\
CH-1015 Lausanne, Switzerland}
\author{H.M.\, R{\o}nnow}
\affiliation{Laboratory for Quantum Magnetism  (LQM), {\'E}cole Polytechnique F\'{e}d\'{e}rale de Lausanne (EPFL), \\ CH-1015 Lausanne, Switzerland}


\maketitle

{\onecolumngrid

{
Magnetic skyrmions are topologically protected  spin-whirl quasiparticles  currently considered as  promising components for ultra-dense memory devices. In the bulk they form lattices that are stable over just a few Kelvin below the ordering temperature. 
This narrow stability range presents a key challenge for applications, and finding ways to tune the SkL stability over a wider phase space is a pressing issue. 
Here we show experimentally that the skyrmion phase in the magnetoelectric insulator $\CuSe$ can either expand or shrink substantially depending on the polarity of a moderate applied electric field. The data are well-described by an expanded mean-field model with fluctuations that show how the electric field provides a direct control of the free energy difference between the skyrmion and the surrounding conical phase. Our finding of the direct electric field control of the skyrmion phase stability offers enormous potential for skyrmionic applications based on a magnetoelectric coupling.


}

}

\newpage

\twocolumngrid

\small

To realise skyrmion-based applications, research into creation, control and stabilisation of skyrmions is in an active phase \cite{Jonietz2010,Seki2012a,Everschor2012,Yu2012,White2012,Fert2013,White2014,Mochizuki2014,Jiang2015}. A clear problem to overcome is that in bulk materials, the skyrmion lattice (SkL) is always only stable over a very narrow range of temperature ($T$) and applied magnetic field ($\mu_{0}H$) just below the critical temperature $T_C$ \cite{Muhlbauer2009,Munzer2010,Adams2012,White2014,Tokunaga2015,Binz2006}. In $\CuSe$ for example, the skyrmion pocket spreads downwards in $T$ by just $3.5 \%$ of $T_C$, occupying no more than $1~\%$ of the total ordered phase space \cite{Seki2012a,Levatic2016}. This generally limited phase space is observed also in other known bulk skyrmion hosts~\cite{Muhlbauer2009,Munzer2010,Wilhelm2011,Tokunaga2015}, and significantly restricts the scope for the development of industrial applications. The ability to enhance or suppress the skyrmion phase space in a sample can provide a flexible platform for the respective creation or destruction of skyrmion states. Here we present a simple and reliable mechanism for the stabilisation and destabilisation of the skyrmion phase as that due to electric ($E$) fields applied to an insulating material.


Up to now, several approaches for skyrmion manipulation were demonstrated using either moderate electric currents, electric fields, or thermal gradients~\cite{Jonietz2010,Yu2012,Everschor2012,White2012,White2014,Mochizuki2014,Jiang2015,Mochizuki2015,Upadhyaya2015,Mochizuki2016,Okamura2016, Kruchkov2017,Watanabe2016}. Progress towards tuning the bulk skyrmion phase stability was also demonstrated using both applied uniaxial \cite{Chacon2015,Nii2015} and hydrostatic pressure \cite{Levatic2016}. For possible applications of the insulating skyrmion host materials, the use of electric field to manipulate the skyrmions is a very promising option that remains still relatively little explored.

Here we report a combined theoretical and experimental study of SkL phase stability under moderate $E$-fields (kV/mm) in the model insulating skyrmion host $\CuSe$. Theoretically the $E$-field effect is addressed using first order perturbation theory around the mean-field solution. This results in a small $E$-field driven shift of the SkL free energy that is nevertheless comparable with the energy difference between the skyrmion and conical phases. Furthermore, we develop a new approach for treating the fluctuative free energy by adding quasiparticle modes near $T_C$ which prove to be pivotal in evaluating the free energy differences between the phases. To verify experimentally the theoretical expectations for the skyrmion phase stability under an $E$-field, we use small-angle neutron scattering (SANS) to study microscopically how the $E$-field controls the extent of the equilibrium skyrmion phase in $\CuSe$. Consistent with our theory, we find that both the magnetic field and temperature extent of the skyrmion phase either expands or shrinks dependent on the $E$-field polarity. In addition, we identify the appropriate experimental conditions for either the enhanced or suppressed skyrmion phase stability in the sample, with similar conditions found in theory by exploring the free energy density map under positive and negative voltages.\\

\no
\textbf{Results}

\no
\textbf{Controlling skyrmion phase stability using electric fields.} From recent bulk susceptibility measurements $(\chi(E))$ of $\CuSe$~\cite{Okamura2016}, it was suggested that skyrmions may be "created" or "annihilated" by applying a dc $E$-field in suitable parts of the $(T,\mu_{0}H)$ phase diagram. In that study~\cite{Okamura2016} the skyrmion phase is identified as a small drop in $\chi(E)$, which serves only as an indirect indication for the existence of the skyrmion phase. Here we use the tool of small-angle neutron scattering (SANS) to directly observe the microscopic skyrmionic magnetism in $\CuSe$, and its response to an applied dc $E$-field. By SANS the skyrmion lattice phase is typically observed as a sixfold symmetric diffraction pattern, consistent with the so-called  multispiral (triple-\textbf{q}) magnetic structure described by three propagation (\textbf{q}-)vectors rotated by 120$^{\circ}$ with respect to each other (note that both $\pm$\textbf{q} each give a Bragg spot)~\cite{Muhlbauer2009,Seki2012,White2014}. To maximise the $E$-field effect, in our SANS experiments we oriented the sample so that $E||\mu_{0}H||[1 1 1]$ and, in comparison with the size of the skyrmion phase for $E$=0, succeeded in observing an expansion (contraction) of the skyrmion pockets for an $E$-field applied parallel (antiparallel) to the $[111]$ axis. The changes in the phase diagram are summarised in Figure~1.


{\onecolumngrid {

%
\begin{figure*}[b]
	\includegraphics[width=0.9 \textwidth]{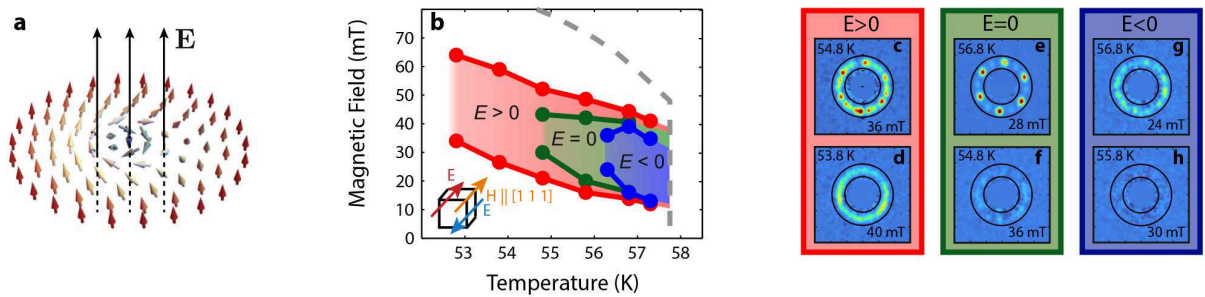}
	\caption{\textbf{Skyrmion phase tuning by electric fields.}
	(a) A sketch of a magnetic skyrmion with the relative direction of the applied electric field $\vec E$, in our experiments $E||\mu_{0}H||[1 1 1]$.
	(b) Phase diagram (skyrmion pockets) measured by small-angle neutron scattering (SANS). The skyrmion pocket spreads almost twice in the positive field $+5.0~\text{kV/mm}$, while shrinking twice under a negative field of $-2.5\text{kV/mm}$.
	(c)-(h) Typical SANS diffraction patterns obtained from the SkL phase under various electric fields. Here the [111] direction is into the page. The black rings define an annular integration window used to evaluate the total scattered intensity due to the skyrmion phase on the detector. The hallmark six-fold symmetric scattering of the skyrmion lattice is clearly observed in (c),(e). At 56.8~K, applying $E<0$ suppresses both the SkL formation (compare (e) and (g)). In panels (d),(f),(g) the six-fold symmetric scattering signal is not clear, but it nevertheless arises from (orientationally-disordered) skyrmion arrays and displays the same $|q|$ as the Bragg spots in six-fold patterns shown in (e) and (g). In panel (h) no SANS signal is observed above the background level.}
	\label{Fig.1}
\end{figure*}
%

}
}

\

\

\

\

\twocolumngrid

\newpage

{\onecolumngrid {

%
\begin{figure*}[t]
	\includegraphics[width=0.85 \textwidth]{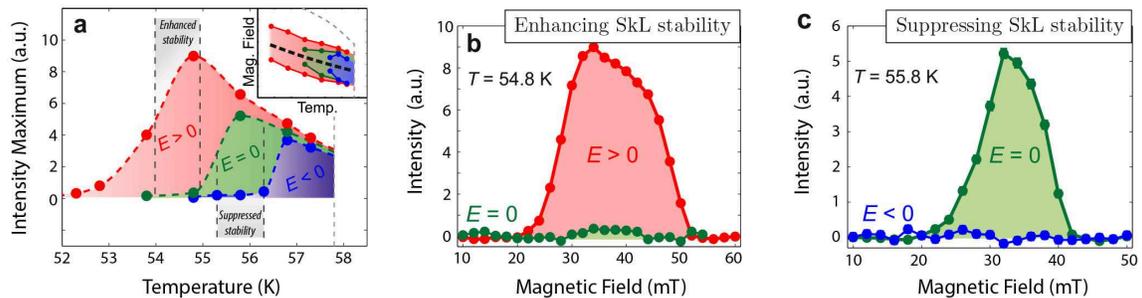}
	\caption{\textbf{Optimising skyrmion stability in electric fields.}
(a) Maximum SANS intensity versus temperature along the direction of the skyrmion pocket growth (dashed line on the inset). The zone favourable for enhancing the skyrmion phase stability ("writing" skyrmions) is 53.9 to 54.9 K, where the skyrmion array population is the highest for $E>0$, while for $E=0$ the skyrmion phase is absent. For suppressing the skyrmion phase stability ("erasing" skyrmions), it is favourable to place the sample between 55.3 to 56.3 K, where the skyrmion phase is well populated under zero voltage, but becomes strongly suppressed under $E<0$.
(b) Total scattered SANS intensity versus temperature for $E=0$ and $E>0$ at $T=54.8 \, \text{K}$  ("stabilising" area). The value of $\mu_{0}H$ at which the SANS intensity is a maximum is plotted in (a). Panel (c) shows similar data as in (b), but for $E=0$ and $E<0$ at $T=55.8 \, \text{K}$  ("destabilising" area).
	}
\end{figure*}
%

}
}

\twocolumngrid

Figure 1 shows representative SANS data collected at various ($T$,$\mu_{0}H$,$E$) conditions, with the sixfold symmetric SANS patterns due to a skyrmion lattice most clearly seen in Figs.~1c,e. In these particular SANS patterns, weaker spots are also detected to lie between the six strongest spots. This indicates the co-existence at various $(T,\mu_{0}H,E)$ conditions of differently oriented skyrmion lattice domains around the $\mu_{0}H$-axis, a phenomenon that has also been reported in other scattering studies of $\CuSe$ \citep{Seki2012,Zhang2016,Makino2016}. For the patterns shown in Figs.~1d,f,g, each obtained near to an edge of the respective skyrmion phase as determined in the SANS experiment, the Bragg spots become ill-defined, and instead the intensity appears as azimuthally smeared patches, indicative of orientationally-disordered SkLs (hereafter termed `skyrmion arrays'). Since the origin of the SkL disordering is difficult to identify unambiguously, a systematic analysis of all SANS data is done by evaluating the the total scattered SANS intensity observed on the detector within the same annular integration window shown in each of Fig.~\ref{Fig.1}b-g. From this approach we account for the scattering due to \emph{all} the skyrmion arrays in the sample when determining the parametric extent of the skyrmion phase.

The main result of the SANS analysis is shown in Figure 1b. Importantly our results show how it is easier to destabilise the skyrmion phase than stabilise it; a positive $E$-field of $+5.0 \text{kV/mm}$ is required to expand the skyrmion pocket so that it becomes almost twice larger, while a negative $E$-field of only $E=-2.5 \text{kV/mm}$ is needed to shrink the pocket by approximately a factor of two. Since this controlled phase expansion and contraction should be expected to occur in general for an insulating skyrmion phase at any temperature, our findings are quite suggestive for applications; for a device layer of thickness $1 \mu m$ (typical for modern electronics) the skyrmion phase in a sample can be almost entirely destabilised (destroyed) or restabilised (restored) with just a couple of volts - the voltage used in a typical smartphone.

\no
\textbf{Optimum conditions for stabilising and destabilising the skyrmion phase}
Examining our SANS analysis more closely shows that at various points in the magnetic phase diagram, the moderate $E$-fields $\text{kV/mm}$ either near-fully destabilise or stabilise the skyrmion phase, when compared with data obtained at the same points but at $E=$~0. As denoted in Fig. 2a, both of these tendencies are observed to occur at a significant level over large $T$ windows each roughly 1~K wide, this corresponding overall to nearly $4\%$ of $T/T_{\rm C}$ in $\CuSe$.

As a representative example, consider the harsh case at when skyrmions are expected to be essentially absent from the system, such as on the conical/Skyrmion phase border at the lower-$T$ boundary of the unperturbed skyrmion pocket ($E=$0). Fig.~2b shows $\mu_{0}H$-scan data obtained for this case at $T=$~54.8~K. While at $E=$0 the lack of SANS intensity indicates the skyrmion phase to be essentially absent, the application of $E=+5.0 \text{kV/mm}$ clearly leads to an enhancement of the skyrmion stability such that significant SANS intensity  of the skyrmion phase is observed. In contrast, the destabilisation of skyrmion arrays requires an $E$-field of opposite sign, and also a slightly higher $T$. Fig.~2c shows that at $T=$~55.8~K, a negative $E=-2.5 \text{kV/mm}$ can completely destabilise the skyrmion arrays that are otherwise stable over a finite range in $\mu_{0}H$ for the unperturbed state ($E$=0).

The optimum $T$ windows for enhancing or suppressing the skyrmion phase stability are labelled in Fig. 2a, and determined from our $\mu_{0}H$-scans of the total scattered SANS intensity obtained at various ($T$,$E$) conditions - see Figs.~2b,c, and also [Supplemental Material]). The total scattered SANS intensity is a quantity indicative of both the population and quality of the skyrmion arrays in the sample. The data in Fig. 2a show the $\mu_{0}H$ at which the SANS intensity is a maximum at each $T$ and $E$-field, and from this the most favourable stability conditions for skyrmion arrays are inferred. We find that the approximate $T$-window of 53.9-54.9~K is appropriate for demonstrating the strongest enhancement of the skyrmion phase stability under $E=+5.0 \text{kV/mm}$, while a $T$-window of 55.3-56.3~K is suitable for demonstrating the suppression of the skyrmion phase stability for $E=-2.5 \text{kV/mm}$. Positioning a device with a good accuracy close to the single  $T$ of 55~K allows to demonstrate both a significant enhancement and suppression of the skyrmion phase by $E$-fields of opposite polarity.\\




\begin{figure*}[]
	\includegraphics[width=0.85 \textwidth]{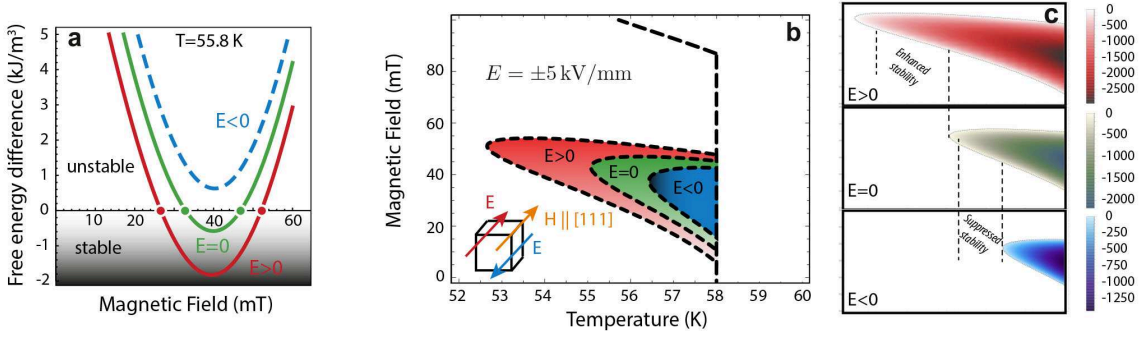}
	\caption{\textbf{Theory: skyrmion lattices in electric fields.} a) Free energy difference between skyrmion and conical phases at $T$=55.8 K for three values of $E$-field ($E=0$,$E=\pm 5 \text{kV/mm}$) in the symmetric-response approach ($\ae^1$). For positive $E$-field (red), the free energy minimum due to the skyrmion phase is deeper than in the absence of voltage (green) meaning that the skyrmion phase is more stable against perturbations; meanwhile the negative voltage destabilises the skyrmion phase (blue dashed line). (b) Phase diagram (skyrmion pockets) for $E$-fields $\pm 5 \text{kV/mm}$, again for $\vec E || \vec H || [1 1 1]$. (c)
	Free energy density map for $E>0$, $E=0$, $E<0$ which represents the skyrmion lattice stability across the phase space. Skyrmion phase stability enhancement and suppression is favourable in the regions with sufficient depth of free energy, as denoted schematically on the plot.}
\end{figure*}



\twocolumngrid

\no
\textbf{Free energy in electric fields.}
The underlying mechanism for either enhancing or suppressing the skyrmion phase stability by $E$-fields is mediated by the magnetoelectric (ME) coupling in insulating $\CuSe$.  Microscopically, the ME coupling originates from the the $d$-$p$ hybridisation mechanism (see Refs. \cite{Jia2007,Belesi2012,Seki2012,Liu2013}). The emergent electric dipole moment $\vec P = \lambda (S_y S_z, S_z S_x, S_x S_y)$ is linked to the underlying spin structure $\vec S (\vec r)= (S_x,S_y,S_z)$ with the coupling parameter $\lambda$ of relativistically small size. Crucially, this effect results in a $\vec P \cdot \vec E$ shift of energy in $E$-field because the skyrmion phase now has a nonvanishing electric-dipole moment. This perturbation renormalises the elementary helices upon which the skyrmion phase is built, and slightly distorts the skyrmion lattice ~\cite{White2014} compared for when $E=$0. 

In this study, we apply the ME perturbation to the free energy described by the effective Ginsburg-Landau functional with Dzyaloshinski-Moriya interaction (DMI), and consider the critical fluctuations which in bulk samples favour the skyrmion phase with respect to the neighbouring conical phase (see Methods). Due to the relativistically small size of $\lambda$, the dimensionless $E$-field is rather small so that, $\ae = \lambda E/ D k_0 \ll 1$, and we can build the perturbation theory in $\ae$ for the modified free energy neglecting all the terms of order $\ae^2$ and higher. Our finding is that the perturbations of fluctuative terms come in only at second order, while the mean-field energy already shifts in the first order due to the direct ME and nonlinear contributions (see Methods). The corresponding shift in free energy of the skyrmion phase depends on the direction of $E$-field (see Fig. 3a), thus either enhancing the skyrmion phase stability ($E>0$) or destabilising ($E<0$) it. While at first sight it can be surprising that perturbatively small $E$-fields play a crucial role here, this is facilitated by the very close competition between the skyrmion and conical phases already in the mean-field.\\

\no
\textbf{Calculation of the phase diagram.} To calculate the phase diagram in $E$-field, we use a new approach developed on the basis of effective models from Refs.\cite{Muhlbauer2009,Janoschek2013,White2014} (see Methods). In contrast to these earlier studies the new approach is more self-consistent in the way that it captures phase diagram, provides a deeper understanding of the role of quasiparticle  modes near $T_C$, and covers the path-integral approach for calculating the fluctuative free energy \cite{Muhlbauer2009} as a limiting case. We thus treat the first-order perturbation in $E$-fields on top of the mean-field solution, and add the fluctuative contributions that stabilise the skyrmion lattice in the bulk. The main contribution to $E$-field effect here is given by the shift of the mean-field free energy difference between the two phases (conical and SkL), while the fluctuative shift under voltages can be considered quadratically small but lead to asymmetrical behaviour as discussed in the next section.

The new approach to the phase diagram calculation allows us to understand deeper the stability of the skyrmion lattice on the intuitive, pictorial level: the critical fluctuations (waves) are superposed on top of the variationally minimised free energies. There are three critical modes $\omega^{(0,1,2)}_{\vec k}$ around the mean-field (see Supplemental Material), with $\omega^{(0)}_{\vec k}$ soft on the sphere $|\vec k|=k_0$, which means that it cost very little energy to add many such fluctuations if they are coherent with the helix $k_0$. Thus $\omega^{(0)}_{\vec k_0}$ is the so-called "dangerous" mode since it results in a Van-Hove-like singularity at $T_C$ and eventually breaks down the ordered phases into the disordered (paramagnetic) phase \cite{Janoschek2013}. Below $T_c$ the breaking of symmetry can be observed by SANS with either a six-fold pattern (skyrmion phase) or two-fold pattern (helical or conical phase), both circumscribed by a sphere $|\vec k|=k_0$ in reciprocal space.  Our calculation shows that the skyrmion phase is favoured  because adding fluctuations costs more entropy in the skyrmion phase. This analysis leads also to a qualitative criterion of the vertical breakdown of the ordered phases at $T_C$ (see Methods). Asymptotically, the main contribution of the fluctuative free energy is given in the short-scale physics, where  $\CuSe$ is "almost" a ferromagnet, thus reproducing the surprising result of the path-integral approach \cite{Muhlbauer2009} as a limiting case. The model described here captures the qualitative physics of the system, as exemplified by the theoretical phase diagram shown in Fig. 3b.

Experimentally, we have observed that the parametric extent (stability) of the skyrmion phase become enhanced under $E>0$. This observation can be addressed theoretically by exploring the free energy density map across the phase space for different values of $E$-fields. We find that the depth of free energy minimum deepens with an increasingly positive $E$-field, as intuitively might be expected. For example, if we sit at $T=54.8 \, \text{K}$ at $E$=0, the free energy of the skyrmion phase has a gap with respect to the conical phase (see Figure 3a), which means that the skyrmion phase is not favoured at this condition; if we now include the $E$-field, there is a finite range of $\mu_{0}H$ where the free energy difference is negative with respect to the conical phase and the skyrmion lattice can now exist. In Fig.~3c we label the regions favourable for either enhancing (stabilising) or suppressing (destabilising) the skyrmion phase stability as spanning approximately half of the calculated $E$-field modified skyrmion pockets.

\

\

\no
\textbf{Discussion}

\no
In some respects, the observed $E$-field effect on the skyrmion phase stability resembles that achieved due to either applied uniaxial~\cite{Chacon2015,Nii2015} or hydrostatic pressure~\cite{Levatic2016}. Clearly however, integrating the pressure effect on skyrmion stability into a technological setting is very challenging. In contrast, the $E$-field effect proves to be both a versatile and reliable external parameter; providing an efficient control of both the skyrmion position~\cite{White2012,White2014,Omrani2014} and the stability of the phase as a whole as demonstrated here.

We also clarify that the underlying mechanisms governing the two phenomena of the pressure and $E$-field effects are different. Namely, in the present $E$-field study, it is a voltage-induced distortion of the SkL which either enhances or suppresses the stability of the skyrmion phase with respect to the conical phase. Similar experimental observations as reported here were recently communicated from indirect bulk susceptibility data without any theoretical support\cite{Okamura2016}. Our present study lays both theoretical and experimental foundations for fully exploring alternative $\mu_{0}H$- and $E$-field configurations, not only in reciprocal-space measurements like SANS, but crucially real-space imaging techniques such as cryo-Lorentz transmission electron microscopy (LTEM). In addition, there is an urgency for studying the $E$-field effect on both equilibrium and metastable skyrmion phases, since each of these can serve as platforms for exploring single-skyrmion creation/annihilation processes, and real-time $E$-field guiding of skyrmions in confined geometries.

In conclusion, we have demonstrated experimentally and understood theoretically the mechanism by which a moderate electric field may either enhance or suppress the stability of the skyrmion phase in the magnetoelectric chiral magnet $\CuSe$. Since our observations can be generally expected to occur in a suitable insulating host material at room temperature, our study provides motivation for the theoretical exploration of skyrmions both over the richer phase space afforded by the $E$-field, and, while awaiting suitable real materials, the development of insulator skyrmion-based data storage and racetrack memory devices where one needs to copy and erase huge arrays of data. In addition, the theoretical approach herein can be extended towards describing the $E$-field effect on stable and metastable skyrmion states in thin films, which are also of paramount technological importance.

\

\no
\textbf{Methods}

\footnotesize{

\no
\textbf{Small-angle neutron scattering (SANS)}. For the SANS experiment, we used a single crystal crystal grown using chemical vapour transport \cite{Belesi2010}. The crystal was of mass 6~mg and volume $3.0\times2.0\times0.50$~mm$^{3}$ with the thinnest axis $\parallel$$[111]$, and $[\bar{1}\bar{1}2]$ vertical. The sample was mounted onto a bespoke sample stick designed for applying dc $E$-fields \cite{Bartowiak2014}. In our experiments we achieved $E$-fields ranging from $+5.0$~kV/mm to $-2.5$~kV/mm. Evidence of electrical breakdown was detected for $E$-fields outside this range.

The sample was loaded into a horizontal field cryomagnet at the SANS-II beamline, SINQ, PSI. The magnetic field ($\mu_{0}H$) was applied parallel to both the $[111]$ direction of the sample and the incident neutron beam to give the experimental geometry $E$$\parallel$$\mu_{0}H$$\parallel$$[111]$. In this geometry, the SANS signal is only detected from the skyrmion phase, which typically presents as a hexagonal scattering pattern with propagation vectors \textbf{q}$\perp$$\mu_{0}H$. In this geometry, we avoid detecting any SANS signal due to either of the neighbouring helical (\textbf{q}$\parallel$$\langle001\rangle$) or conical phases (\textbf{q}$\parallel$$\mu_{0}H$), since the propagation vectors of these phases lie well out of the SANS detector plane.

We used incident neutrons with a wavelength of 10.8~\AA ($\Delta\lambda/\lambda=10\%$). The scattered neutrons were detected using a position-sensitive multidetector. The SANS measurements were done by rotating (`rocking') the sample and cryomagnet ensemble over angles that brought the various SkL diffraction spots onto the Bragg condition at the detector. Data taken at 70~K in the paramagnetic state were used for background subtraction. Before starting each $\mu_{0}H$-scan, the sample was initially zero field-cooled from 70~K to a target temperature, with the $E$-field applied when thermal equilibrium was achieved. The $E$-field was maintained during the $\mu_{0}H$-scan. At each $T$ we define the $\mu_{0}H$ extent of the SkL phase as that over which SANS intensity is detected. We use this criterion to extract the parametric extent of the SkL phase for ($\mu_{0}H$,$T$,$E$) as shown in Figures 1,2. See Supplemental Material for more details.

\

\no
\textbf{Mean-field free energy.}
The effective mean-field theory is based on the coarse-grained magnetisation approach $M(\vec r) = M_s \vec S(\vec r)$ and is sufficiently described in \cite{Muhlbauer2009}. One starts with the mean-field approach with free energy

\begin{align}
\label{free energy}
F[\vec M] =  \langle   \Theta_T \, \vec M^2 + J (\nabla \vec M)^2 +  D \, \vec M \cdot (\nabla \times \vec M)
  + U \vec M^4   -  \vec H \cdot \vec M
\rangle
\end{align}

\no
where the averaging is  $\langle... \rangle = \int {\frac{dV}{V}}...$, and $\Theta_T \propto \alpha (T-T_C)$ near $T_C$, $J$ is the Heisenberg stiffness and $D$ is DMI, $H$ is the magnetic field, and the higher-order term  $U$  grants the formation of the crystalline phase \cite{Muhlbauer2009}. In the mean-field, the interplay between Heisenberg and DMI energies determines the helical vector as $k_0 = D/2 J$. The long-range-ordered hexagonal skyrmion lattice is approximated as
$ \vec S (\ra)  \simeq \vec m
+ \mu
\sum_{\vec q_n}  \vec S_{\vec q_n} e ^{i \vec q_n \ra+i \varphi_n}
+  \text{c.c.}$,
where the summation runs over the crystalline order vectors $\vec q_1+\vec q_2+\vec q_3=0$. In the mean-field, the skyrmion phase is slightly gapped with respect to the conical phase, however the two are closely competing. Further details of the mean-field theory described in Ref. \cite{Muhlbauer2009}.

\

\no
\textbf{Perturbation theory in electric fields.}
The magneto-electric coupling in $\CuSe$ is relativistically small, so the perturbation parameter is $\ae = \lambda E/4 D k_0 \ll 1$. It is sufficient to use the first order perturbation theory on top of the non-perturbed free energy.
We go to the rotated frame defined by the magnetic field direction along [1 1 1], and re-write the free energy.  The first order perturbation theory gives eigenvectors:

\begin{equation} \begin{aligned}
\ket{ {\vec S}^{(\ae)}_{\vec k}  }
 =
\ket{ {\vec S}^{(0)}_{\vec k} }
+\sum_{n \ne 0}
\ket{ {\vec S}^{(n)}_{\vec k}  }
\frac{
\bra{ {\vec S}^{(n)}_{\vec k}  }
\hat {\mathcal{H}}_{E}
\ket{ {\vec S}^{(0)}_{\vec k} }
}
{\varepsilon^{(0)}_{\vec k}-\varepsilon^{(n)}_{\vec k}} + \mathcal{O} (\ae ^2) ,
\end{aligned}\end{equation}

\no
which are now the basis for constructing the distorted skyrmion lattice. For other field ($\vec H$, $\vec E$) configurations, we re-do the calculations in the new rotated frames. See Supplemental Materials for further details.

\

\no
\textbf{Fluctuation-induced phase stabilisation.}
We use a new approach, which captures as a limiting case the fluctuation free energy from \cite{Muhlbauer2009}.
The essential physics is captured already in Gaussian (noninteracting) fluctuations with free energy density

\begin{equation}\begin{split}\begin{gathered}
\label{fluctuations}
F_{\text{fluct}}
 =
 \sum_{i} \sum^{|\ka|<\Lambda}_{\ka}
\omega_{\ka}^{(i)} f^{(i)}_{\ka}  - T \, \text{S}_{\text{fluct}},
\end{gathered}\end{split}\end{equation}

\no
where $\Lambda = 2 \pi /a $ is the natural cut-off, $f^{(i)}_{\ka}$ is the critical modes distribution, and the entropy of Gaussian fluctuations is

\begin{equation}\begin{split}\begin{gathered}
\label{entropy}
\text{S}_{\text{fluct}}
 =
 \sum_{i} \sum^{|\ka|<\Lambda}_{\ka}
\{
(1+ f^{(i)}_{\ka} ) \, \ln(1 + f^{(i)}_{\ka} )
-
f^{(i)}_{\ka}  \ln f^{(i)}_{\ka}
\}
\end{gathered}\end{split}\end{equation}

\no
in the case of bosons.
Fluctuations around mean-field are described by the generalised susceptibility $\chi^{-1}_{i j} (\vec r, \vec r ')  = \frac{1}{T}
\frac{\delta^2 F}{\delta M_i (\vec r) \, \delta M_j (\vec r') }$, giving rise to several collective modes
(See Supplemental Material).
On the local scale (or $k \gg J/D$), the chiral magnet is reminiscent of a ferromagnet, so the modes behave asymptotically $\omega_{\ka} \propto k^2$ for large $k$, thus asymptotically  $F_{\text{fluct}}   \simeq \log{\beta \omega_{\ka}} \propto \log k^2$, which covers the model of Ref.\cite{Muhlbauer2009}. The main contribution to \eqref{fluct energy} is given by the short length-scale ("ferromagnetic") physics,

\begin{equation}
\begin{aligned}
\label{fluct energy}
 \Delta F_{\text{fluct}}  \simeq
 \frac{10 U}{\pi a J}
\langle \vec S^2_{\text{SkL}} - \vec S^2_{\text{con}} \rangle T,
\end{aligned}
\end{equation}

\no
The electric field also slightly affects the fluctuative energy, because it modifies the correlation length near $T_C$ and so renormalises $J_{\text{eff}}$, which is neglected here as a higher-order ($\ae^2$) effect. See [Supplemental Material] for further details.

\

\textbf{Parameters of the effective model.} For our numerical calculations we use $T_C =58 \, \text{K}$, which approximately sets the Heisenberg stiffness as $J = 4.85 \times 10^{-23} \, \text{Jm}/\text{A}^2$. From the SANS measurement we establish directly the modulation period of $60 \, \text{nm}$, which estimatively differs by a few percents from the mean-field value $2 \pi/k_0$, because the mean-field ordering vector $k_0=D/2J$ is slightly renormalised by the fluctuations near $T_C$. This sets the "bare" DM interaction entering \eqref{free energy} as $D=-9.85 \times 10^{-15} \, \text{J}/\text{A}^2$. The lattice parameter is $a=8.91 \times 10^{-10} \, \text{m}$, which gives the natural cutoff $\Lambda = 2 \pi /a \approx 70 \, k_0$. The saturation magnetization in $\CuSe$ is $M_s = 1.11 \times 10^5 \, \text{A/m}$ and scales with temperature as $M_s(T) = M_s (1- (T/T_C)^{\alpha_1})^{\alpha_2}$, with $\alpha_1= 1.95$ and $\alpha_2 = 0.393$ \cite{Zivcovic2012}. We choose the nonlinear coupling responsible for SkL formation $K = 6.2 \times 10^{-6} \, \text{J} \text{m}^{-1}\text{A}^{-2}$ and Landau parameter $\alpha_T = \theta_T / J k_0^2 (T-T_C)= 3.5 \, \text{K}^{-1}$. For the qualitative phase diagram shown in Fig. 3b, we use a symmetric-response model ($\ae^1$), for which the best fit to SANS data is for $\ae = 0.02$, which corresponds here to $E=\pm 5 \times 10^6 \, \text{V}/\text{m}$  coupled with $\lambda/D k_0 = 9.23 \times 10^{-9} \, \text{m}/\text{V}$ to the underlying spin structure through ME mechanism.

\

\textbf{ACKNOWLEDGEMENTS}

 We thank  Achim Rosch, Naoto Nagaosa and Jiadong Zang for useful discussions. The work was supported by the Swiss National Science Foundation, its Sinergia network Mott Physics Beyond the Heisenberg Model (MPBH). Neutron scattering experiments were carried out at the Swiss Spallation Neutron Source (SINQ), Paul Scherrer Institut, Switzerland.

\end{document}